\documentclass[aps,pra,twocolumn,final,floatfix,superscriptaddress,10pt]{revtex4-1}
\usepackage[utf8]{inputenc}
\usepackage{physics}
\usepackage{mathtools}
\usepackage{amsmath}
\usepackage{xcolor}
\usepackage{tikz}
\usepackage{qcircuit}
\usepackage{soul}

\usetikzlibrary{graphs,graphs.standard}

\begin{document}

\title{Approximating the quantum approximate optimization algorithm with digital-analog interactions}
\author{David Headley} \email{David.Headley@mercedes-benz.com} \affiliation{Mercedes-Benz AG, Stuttgart, Germany}  \affiliation{Theoretical Physics, Saarland University, 66123 Saarbr{\"u}cken, Germany}

\author{Thorge M\"uller} \email{Thorge.Mueller@dlr.de}
\affiliation{German Aerospace Center (DLR), 51147 Cologne, Germany}\affiliation{Theoretical Physics, Saarland University, 66123 Saarbr{\"u}cken, Germany}

\author{Ana Martin} \affiliation{Department of Physical Chemistry, University of the Basque Country UPV/EHU, Apartado 644, 48080 Bilbao, Spain}\affiliation{EHU Quantum Center, University of the Basque Country UPV/EHU, Bilbao, Spain}

\author{Enrique~Solano} \affiliation{Kipu Quantum, 10405 Berlin, Germany} \affiliation{IKERBASQUE, Basque Foundation for Science, Plaza Euskadi 5, 48009 Bilbao, Spain}
 \affiliation{International Center of Quantum Artificial Intelligence for Science and Technology (QuArtist) \\ and Department of Physics, Shanghai University, 200444 Shanghai, China}

\author{Mikel Sanz} \affiliation{Department of Physical Chemistry, University of the Basque Country UPV/EHU, Apartado 644, 48080 Bilbao, Spain}\affiliation{EHU Quantum Center, University of the Basque Country UPV/EHU, Bilbao, Spain}
\affiliation{IKERBASQUE, Basque Foundation for Science, Plaza Euskadi 5, 48009 Bilbao, Spain}
\affiliation{Basque Center for Applied Mathematics (BCAM), Alameda de Mazarredo 14, 48009 Bilbao, Basque Country, Spain}

\author{Frank K. Wilhelm} \affiliation{Theoretical Physics, Saarland University, 66123 Saarbr{\"u}cken, Germany}\affiliation{Institute for Quantum Computing Analytics (PGI 12), Forschungszentrum J\"ulich, 52425 J\"ulich, Germany}
\date{\today}

\begin{abstract}
    The Quantum Approximate Optimisation Algorithm was proposed as a heuristic method for solving combinatorial optimisation problems on near-term quantum computers and may be among the first algorithms to perform useful computations in the post-supremacy, noisy, intermediate scale era of quantum computing. In this work, we exploit the recently proposed digital-analog quantum computation paradigm, in which the versatility of programmable universal quantum computers and the error resilience of quantum simulators are combined to improve platforms for quantum computation. We show that the digital-analog paradigm is suited to the quantum approximate optimisation algorithm due to the algorithm's variational resilience against the coherent errors introduced by the scheme. By performing large-scale simulations and providing analytical bounds for its performance in devices with finite single-qubit operation time we observe regimes of single-qubit operation speed in which the considered variational algorithm provides a significant improvement over non-variational counterparts in the digital analog scheme.
\end{abstract}

\maketitle

\section{Introduction}

Quantum computing is entering an era in which classical computers cannot simulate the behaviour of programmable quantum computers \cite{arute2019quantum}. In this new era of quantum information processing, it is likely that the first algorithms that will be useful for solving computational problems will be \textit{heuristic} in nature. These algorithms come without provable performance guarantees provided by the likes of Shor's factoring algorithm \cite{shor1994algorithms} or the Grover search algorithm \cite{grover1996fast}, but are encouraged by strong motivation from classical algorithm research, in that the most effective algorithms for solving certain problems classically are often not provably so. At present, there are two such algorithms that are most likely to prove useful in the near term \cite{Preskill_2018}---The Variational Quantum Eigensolver (VQE) \cite{peruzzo2014variational} and the Quantum Approximate Optimisation Algorithm (QAOA) \cite{farhi2014quantum} otherwise known as the Quantum Alternating Operator Ansatz \cite{hadfield2019quantum}. These are \textit{variational} algorithms, using classical optimiser and parameterised quantum circuits to mitigate the effects that errors may introduce on quantum devices making no use of quantum error correction. This work concerns the latter of the two.

QAOA is a discrete-time hybrid quantum-classical algorithm for computing solutions to problems in combinatorial optimisation. The algorithm was initially discovered to provide greater approximation ratios than the best known classical algorithm for the problem type MAX-E3LIN2 \cite{farhi2014quantum2}, a result later ceded to a quantum-inspired classical algorithm \cite{barak2015beating}. It was demonstrated by Jiang et al. \cite{jiang2017near} that QAOA can recover the square root scaling of the Grover's search algorithm, replicating Grover's speed-up without the need for Grover's mixing operator. Hadfield et al. discovered that QAOA driving operators can be modified such that a wide variety of problems can be solved without resorting to high-order penalty-terms usually considered in an annealing or adiabatic-based approach \cite{hadfield2019quantum}.

The development of QAOA was motivated by a need for algorithms that can run on noisy, pre-error correction devices. Algorithms used on devices of this era will necessarily have a degree of co-design between architecture and algorithm. Work by Rigetti, for example, used a noisy, programmable quantum device to solve a combinatorial optimisation problem inspired by the on-device layout of qubits \cite{otterbach2017unsupervised}. Following this approach, we extend the work of Parra-Rodriguez et al. on the Digital Analog (DA) paradigm of quantum computation \cite{lamata2018digital, parra2020digital, martin2020digital}, in which a device is designed and operated in the style of a \textit{quantum simulator} with always-on multi-qubit interactions. We show that QAOA is a natural algorithm for this setting. This paradigm, leveraged to minimise errors associated with turning on and off gates on a quantum device, could allow for a simpler design in which only the timing of single-qubit gates must be considered, reducing the control complexity and, therefore, the mechanisms through which environmental noise can corrupt the computing system. The problem of negative interaction times, introduced by the aforementioned scheme, is resolved in this work by exploiting periodicity in time applied of the resource interaction, or problem solved, allowing any two-local problem to be solved with a homogeneous resource interaction and problems exhibiting periodicity (MAX-CUT, MAX-2-SAT) to be solved with heterogeneous resource interactions.

The paper begins with a mathematical description of digital-analog computational paradigm, QAOA, and the combination thereof. The costs of compilation for embedding QAOA within the Digital Analog paradigm are examined and the potential hardware platforms on which digital analog quantum computing for this purpose could be performed are discussed. The performance of DA-QAOA is examined using computational simulations and the errors of the method are analytically bounded. It is finally concluded that while using the Digital Analog scheme for QAOA results in additional errors, these errors are not as damaging as in other potential uses of the digital analog paradigm due to the variational nature of QAOA. The novel result of this paper is that the combination of DA computation with QAOA is synergistic. For realistic device parameters, errors introduced by DA computation do not adversely affect the performance of QAOA.

\section{The Digital-Analog Quantum Computational Paradigm}
\label{section-DASCHEME}

\begin{figure}
    \centering
    \includegraphics[scale =0.8]{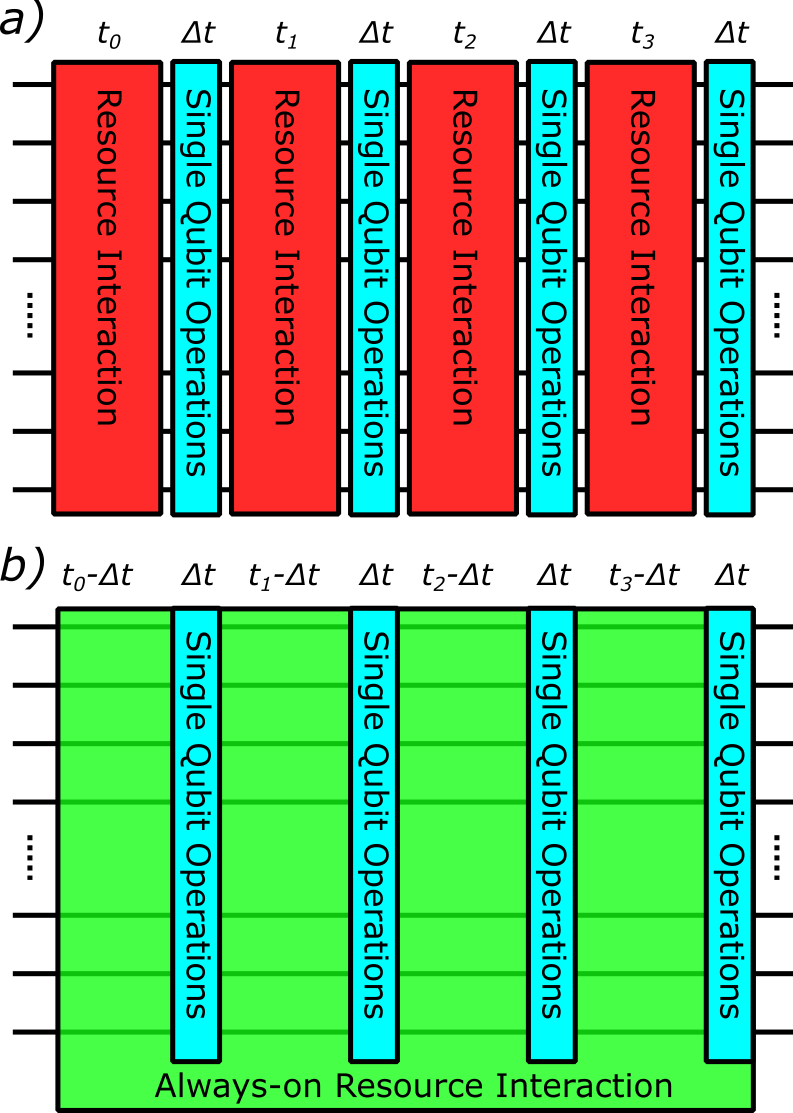}
    \caption{The two schemes for digital analog computation. a) The stepwise or sDAQC scheme in which a series of programmable \textit{digital} single qubit gates are applied in alternation with \textit{analog} resource interactions. b) The always-on or bDAQC scheme in which the resource interaction is never turned off and single qubit operations are applied in parallel with the resource interactions. Performing the single qubit operations simultaneously with the resource interaction introduces coherent errors but reduces device control requirements. The first interaction block denoted with the time interval $t_0$ corresponds to the \textit{idle} block.}
    \label{fig:DA_scheme}
\end{figure}
\begin{figure}
    \centering
    \hspace{-0.5cm}
    \includegraphics[scale = 1.]{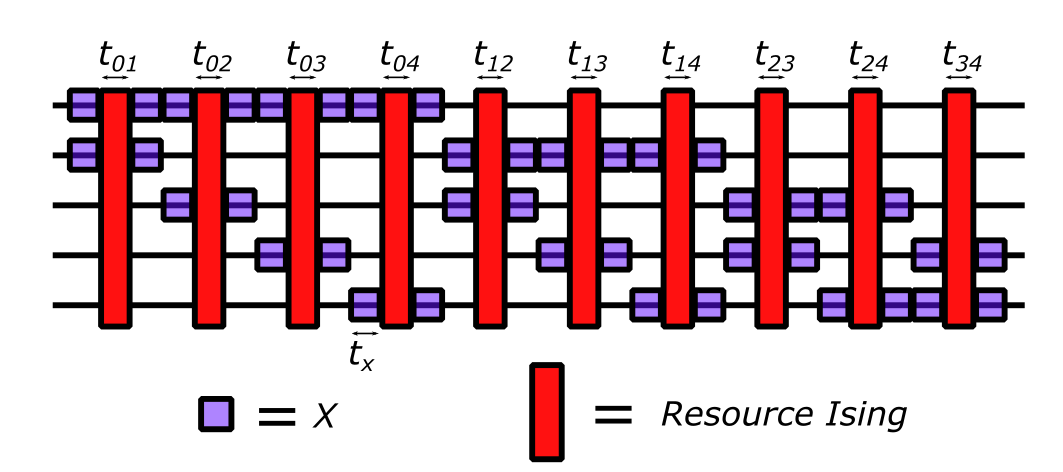}
    \vspace{-0.5cm}
    \caption{A quantum circuit depicting the digital-analog time evolution required to simulate an arbitrary Ising Hamiltonian on $5$ qubits. $10$ uses of a resource Hamiltonian are required, each surrounded by a unique combination of single-qubit-$X$ operations.}
    \label{fig:da_n=5}
\end{figure}

Quantum algorithms can, in general, be separated into two classes: continuous and discrete. At the extreme end of continuous quantum algorithms lie those of quantum simulators, devices fabricated to follow dynamics of interest with, however, no capacity for complicated programmed time evolution \cite{RevModPhys.86.153}. Likewise, continuous algorithms such as the quantum adiabatic algorithm \cite{farhi2000quantum, albash2018adiabatic} and quantum random walk algorithms \cite{kempe2003quantum, kendon2006random} make use of a predefined \textit{analog} Hamiltonian with generally limited programmability to solve computational problems. Discrete algorithms, on the other hand, are defined in terms of sequences of \textit{digital} unitary gates. A gate-based quantum computer can perform any discrete, gate-based algorithm, including Trotterized versions of continuous algorithms \cite{lloyd1996universal}. Whereas, a device designed for continuous quantum algorithms, though may be theoretically universal, will generally only be able to run the restricted set of time evolutions for which they are built---it is in theory possible to express any discrete time quantum algorithm (e.g. Shor's) as a continuous time quantum algorithm, but generally not practical \cite{aharonov2004adiabatic}. Devices of analog quantum computation such as quantum simulators can benefit from superior noise resilience characteristics stemming from a reduced requirement to completely control the full dynamics of every qubit \cite{Ac_n_2018}, as is required in a fully gate based, digital model.

The DA paradigm \cite{lamata2018digital, parra2020digital} is designed to take the best features of both digital and analog quantum computing and has been shown to yield an implementation for the quantum Fourier transform and the Harrow-Hassidim-Lloyd algorithm, in which significant advantages over regular digital schemes are demonstrated for reasonable coherent-control error assumptions \cite{martin2020digital, martin2022digital}. The basic premise of the DA scheme is that blocks resembling the time evolution of an analog simulator are performed, punctuated by digital single-qubit operations. This premise is depicted in quantum circuit form in Fig. \ref{fig:DA_scheme}. The DA paradigm yields two options for its implementation, the \textit{step-wise} scheme (sDAQC) and the \textit{always-on} or \textit{banged} scheme (bDAQC), in which error is introduced to computations due to the non-commutativity of single qubit operations used simultaneously with an entangling resource interaction. This work proposes the use of an always-on-resource (bDAQC-QAOA) on near-term quantum devices for discrete optimisation. To this aim, the step-wise scheme provides a guarantee that if single-qubit gates can be applied sufficiently fast, no errors are introduced and the always on scheme produces the same state as a standard gate-based implementation.

We consider a context in which we have access to a \textit{resource} Hamiltonian consisting of a sum of all possible interaction $ZZ$ terms between connected qubits $j,k$ on an $n$-qubit device, each with relative strength $r_{jk}$,
\begin{equation}
    H_\textrm{Resource} = \sum_{j<k}^n r_{jk}Z_jZ_k.
\end{equation}
Where sums in this work over qubit indices start from $1$. We use this resource as an interacting operation to simulate problem Hamiltonians required for QAOA. In sDAQC we assume that our device has the capability for this resource Hamiltonian to be turned on and off, alongside the ability to perform arbitrary single-qubit gates. We can, therefore, alternate our resource interaction with single-qubit gates. In bDAQC we do not assume resource Hamiltonian can be turned off and on, but retain the ability to perform arbitrary single-qubit gates at any time, simultaneously applied alongside the resource interaction.

Let us assume that the resource Hamiltonian available is an all-to-all (ATA) Ising Hamiltonian:
where in the \textit{homogeneous} case, $r_{jk} = 1\,\, \forall\,\, j,k$ (the units of $r$ are set such that $1$ represents a relevant energy scale for the device). In the simulations and resource estimates performed in this work, we consider only resource Hamiltonians in which no element of $r$ is zero. However, the following procedure does not require an all-to-all connected resource interaction to succeed. Here, we require only that the interactions present in the arbitrary Hamiltonian that we intend to simulate are also non-zero in the resource Hamiltonian. Although this is not necessary in general \cite{galicia2020enhanced}, we assume it for the sake of simplicity. Experimental settings likely to provide such resource Hamiltonians are discussed in section \ref{hardware}.

To use a fixed interaction resource Hamiltonian to simulate the time evolution of an arbitrary spin glass Hamiltonian, older techniques developed for quantum computing with Nuclear Magnetic Resonance spin systems can be used. Named average Hamiltonian theory \cite{cory2000nmr,viola1999universal}, one can design a sequence of interactions and single-qubit operations such that the time average of such evolutions is identical to that of a Hamiltonian of interest.

There are $n(n-1)/2$ individual degrees of freedom in an arbitrary Hamiltonian to be simulated by our resource. We will, therefore, require $n(n-1)/2$ time intervals over which the resource is applied, each with surrounding single-qubit operations such that each block is linearly independent of the others. In order to select the single-qubit operations with which to surround uses of the resource interaction, one can pick the $n(n-1)/2$ ways one can select two of $n$ qubits, applying $X$ gates to these qubits before and after an application of the resource Hamiltonian \cite{parra2020digital}. This choice of block-surrounding operations can be seen in figure \ref{fig:da_n=5} for $n=5$ qubits. Between adjacent blocks, single-qubit-$X$ operators will sometimes cancel, reducing the total number of single-qubit gates required substantially.

The unitary evolution we wish to implement is that of an Ising Hamiltonian with arbitrary couplings:  
\begin{equation}
    U_{\textrm{Arb}}(t) = e^{iH_{\rm Arb}t} \quad \textrm{and} \quad H_\textrm{Arb} = \sum_{j<k}^ng_{jk}Z_j Z_k,
\label{defproblem}
\end{equation}
which we wish to express as a sequence of digital analog blocks. One notes that the time evolution implemented by a digital analog block with $X$ operators on qubits $a,b$ is equivalent that of a constant effective Hamiltonian $X_aX_b H_{Res} X_aX_b$ as
\begin{equation}
(X_a\otimes X_b) e^{t_{ab}H_\textrm{Res}} (X_a\otimes X_b) = e^{t_{ab}(X_a\otimes X_b)H_{Res} (X_a\otimes X_b)}.
\label{none}
\end{equation}
This immediately follows from the identity:
\begin{equation}
e^{itUVU^{\dagger}}= \sum_{k=0}^\infty \frac{(it)^k(UVU^{\dagger})^k}{k!} = Ue^{itV}U^{\dagger},
\end{equation}
valid for any unitary operator $U$. Using the above, we may write an arbitrary Ising Hamiltonian as a sum of digital analog effective Hamiltonians, one for each of $n(n-1)/2$ blocks.
\begin{equation}
    H_\textrm{Arb} = \sum_{j<k}^n\sum_{l<m}^nt_{lm} r_{jk} X_lX_mZ_jZ_kX_lX_m, \label{DA-ops-HAM}
\end{equation}
    for some vector of times $\vec t$ to be computed. An illustration of this Hamiltonian applied on a $5$-qubit device can be seen in figure \ref{fig:da_n=5}. Using the identity $X_iZ_i \equiv -Z_iX_i$ to commute Pauli-$X$ operators to cancellation one obtains:
\begin{equation}
    \sum_{j<k}^n\sum_{l<m}^nt_{lm}r_{jk} (-1)^{\delta_{lj} +\delta_{lk}+ \delta_{mj} + \delta_{mk}} Z_jZ_k.
\end{equation}
Through this expression we replace $n(n-1)/2$ possible interaction strengths $g_{jk}$ between qubits $j,k$ with $n(n-1)/2$ resource interaction times $t_{lm}$ sandwiched by single-qubit-$X$ operators on qubits $l,m$. Using the linear independence of Pauli strings, we can write
\begin{equation}
    \frac{g_{jk}}{r_{jk}} = \sum_{l<m}^n t_{lm}(-1)^{\delta_{lj} +\delta_{lk}+ \delta_{mj} + \delta_{mk}}
\end{equation}
in which finding $g_{jk}$ is a matrix inversion problem made apparent by consolidating the parameter pairs $l,m$ and $j,k$ each to one parameter 
\begin{equation}
\kappa = n(l-1) - \frac{l(l+1)}{2} + m,
\end{equation}
\begin{equation} 
\mu = n(j-1) - \frac{j(j+1)}{2} + k. 
\end{equation}
We arrive at a solution time vector in time scaling at most $O(n^6)$ using Gaussian elimination on a classical computer of a matrix with dimension $n(n-1)/2\times n(n-1)/2$ of
\begin{equation}
t_\kappa = M^{-1}_{\kappa\mu}(g/r)_\mu  \quad \mathrm{for}\quad M_{\kappa\mu} = (-1)^{\delta_{lj} +\delta_{lk}+ \delta_{mj} + \delta_{mk}}.
\end{equation}
For the case of $n=4$, $M$ is singular as, for example, $X_1X_2 H_{resource}X_1X_2 = X_3X_4 H_{resource}X_3X_4$ and the condition of linear independence of the effective Hamiltonians from different blocks is not met. This, however, is a special case and the obtained Hamiltonians are linearly independent for $n>4$. QAOA problems of interest, however, far exceed this value in size.

An obstacle for the usage of this scheme is that any of the times calculated in this procedure may be negative. Following the computation of a time vector $t_\kappa$ providing a DA circuit to simulate a desired Ising Hamiltonian, negative times must be eliminated as it is experimentally impossible to run an always-on interaction---the nature of which we can't temporarily change---for a negative time. 

\section{Negative Digital-Analog Block Times}

In this section, a procedure is presented that exploits the case in which the resource Hamiltonian is homogeneous, or one of $\vec g,\vec r$ takes only values with some high least common multiple, resulting in periodic time-evolution. This condition holds for MAX-CUT and SAT problems considered in this work (all $ZZ$ Hamiltonian terms in section \ref{DAQAOA} have at least half-integer pre-factors or integer multiples thereof). In the case that the resource Hamiltonian is homogeneous, any negative time-block can simply be run for a positive time $2\pi + t$, exploiting the periodicity of the unitary effected as $e^{iHt} = e^{iH(2\pi +t)}$, for $r_{jk} = 1 \,\,\forall\,\, j,k$. As such, with homogeneous resource Hamiltonians we can always replace $t_{lm}$ with $t_{lm} \mod 2\pi$. This technique, involving a homogeneous resource, is unfortunately undesirable as a method to rectify all negative time blocks as we will take a time interval that is typically small and replace it with a larger time $2\pi - |t_{lm}|$. This will result in a DA schedule of single-qubit gates and analog block times requiring an longer total time to run on hardware, incurring greater error rates.

For an inhomogeneous resource Hamiltonian we need to consider one additional time-block surrounded by no single-qubit operations. To determine the size of this \textit{idle} block required, consider
\begin{equation}
     M\vec t = M (\vec t - t_{\textrm{min}}\vec 1 + t_{\textrm{min}} \vec 1).
\end{equation}
$M$ admits $\vec 1$ as an eigenvector with eigenvalue $\lambda$. Intuitively, $\vec 1$ is an eigenvector of $M$  because when applying all possible two-$X$-surrounded DA blocks for an equal time, the time evolutions mostly cancel out leaving a smaller but homogeneous effective interaction. This produces
\begin{equation}
   M\vec t = M(\vec t - t_{\textrm{min}}\vec 1) + \lambda t_{\textrm{min}} \vec 1 \\
\end{equation}
and considering a new, non-negative time vector $\vec t^* = \vec{t} - t_{\textrm{min}}\vec1$
\begin{equation}
   M\vec t = M \vec t^* + \lambda t_{\textrm{min}} \vec 1.
\end{equation}
Applying all possible two-qubit-$X$ DA blocks does not, however, result in a similar homogeneous contribution to the simulated Hamiltonian to resource Hamiltonian ratio for all system sizes. We wish to have an eigenvalue $\lambda$ that is negative, such that when multiplied by negative $t_\textrm{min}$ we produce a positive idle time. Unfortunately, the contributions to the ratio for NISQ-relevant cases with $n>6$ are, themselves, positive. The relation between $n(n+1)/2 + 1$ time-intervals and the Hamiltonian simulated can be written as
\begin{equation}
g_\kappa =  M_{\kappa\mu}t^*_\mu r_\kappa + t_{\rm idle} r_\kappa \label{fullDAequation}
\end{equation}
with $t_{\textrm{idle}} = \lambda t_\textrm{min}$. We solve the negativity problem by letting the always-on resource Hamiltonian run for time $\lambda t_\textrm{min}$ surrounded by no single-qubit gates if $t_\textrm{min}$ is negative. Since $t_\textrm{idle}$ is negative for relevant cases of $n>6$, we must use one of two methods to change the sign of this time, depending on whether a homogeneous or inhomogeneous resource Hamiltonian is available. If the resource is homogeneous we can evolve for time $t_\textrm{idle} \mod{ 2\pi}$, as before. This cost of running for this positive time will only add a small contribution to the total algorithm run-time since it only occurs once per set of DA blocks. In realistic experimental cases, however, we expect only non-homogeneous resource Hamiltonians to be available. Even with non-homogeneous resource Hamiltonians, non-negative idle time is still possible through exploiting properties of the simulated problem Hamiltonian. By setting $H_{\textrm{Arb}} \to -H_{\textrm{Arb}}$ in equation (\ref{defproblem}) and using the fact that all $ZZ$ coupling constants in $H_{\textrm{Arb}}$ will be integer multiples of $1/2$ or zero for MAX-CUT and MAX-2-SAT problems, we can simulate the Hamiltonian of correct sign by exploiting the periodicity of the unitary effected, as $e^{itH_{\textrm{Problem}}} =e^{i(-t)(-H_{\textrm{Problem}})} =  e^{i(-t \mod{ 2\pi})(-H_{\textrm{Problem}})}$. This factor of $-1$ in front of the problem Hamiltonian can then be absorbed into the matrix $M$ in equation (\ref{fullDAequation}) causing the eigenvalue $\lambda$ to become negative. This allows a positive idle-time correction, resolving the negative sign issue for inhomogeneous cases.

The method presented here provides a convenient decomposition of an arbitrary Ising Hamiltonian into time-blocks of our resource interaction surrounded by two pairs of single-qubit rotations. The problem of negative times is resolved for the case of resource or target Hamiltonians satisfying certain constraints. For Hamiltonians not satisfying the aforementioned constraints, approaches including the decomposition into multiple DA sequences satisfying these constraints, or a strategy involving a higher number of analog blocks could still be pursued. To incorporate a greater number of time blocks one could solve the under-determined linear system $\vec g = M\vec t $ with $\vec t \geq 0$ where the time-vector $\vec t$ is of higher dimension than $\vec g$ and, therefore, $M$ is no longer square (and invertible). Approaches to solve such a problem are complicated by the time non-negativity constraints and require a quadratic programming approach. Recent work by Galicia et al. \cite{galicia2020enhanced} extends the digital-analog paradigm to the scenario in which only interactions available on a device with linear, nearest-neighbour connectivity can be used to systematically produce an all-to-all connected arbitrary Hamiltonian. Strategies for architectures that are more connected than linear, yet not fully connected, can therefore also produce arbitrary Ising Hamiltonians, by restriction to a linear chain, or by manually inserting the \textsc{swap} operations of a \textsc{swap} network \cite{ogorman2019generalized}, themselves compiled to digital-analog sequences.

\section{The Quantum Approximate Optimisation Algorithm}

\begin{figure}
    \centering
    \includegraphics[scale =0.8]{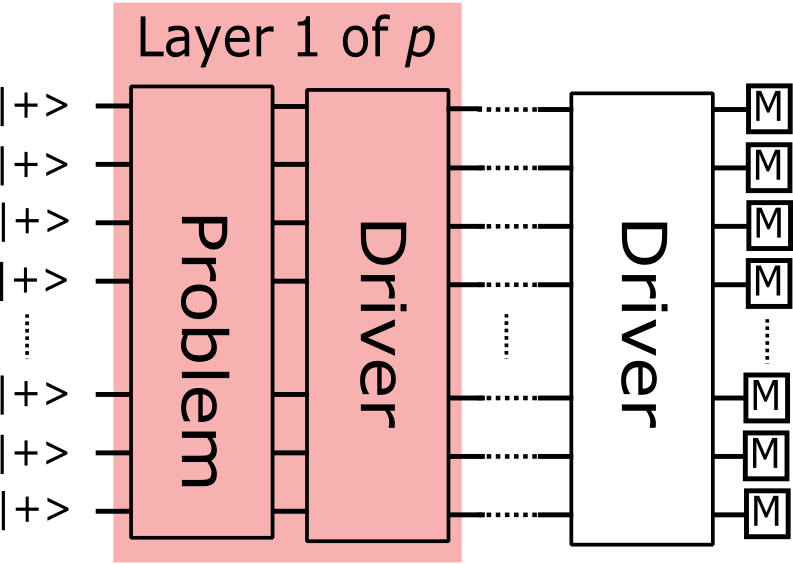}
    \caption{A typical QAOA state preparation circuit. A problem and driver Hamiltonian are alternated $p$ times, applied to the $\ket+^n$ state, followed by measurement on all qubits.}
    \label{fig:QAOA}
\end{figure}

The Quantum Approximate Optimisation Algorithm is a hybrid quantum-classical algorithm in which a classical optimiser tunes $2p$ parameters $\vec \gamma, \vec \beta$ of a quantum circuit to maximise the objective function of a combinatorial optimisation problem. In QAOA, an ansatz state 
\begin{equation}
\ket{\vec \beta,\vec\gamma}  =\prod_{p'=1}^p e^{i\beta_{p'}H_{\rm D}} e^{i\gamma_{p'} H_{\rm P}}\ket+^{\otimes n} \label{QAOA_state}
\end{equation}
is generated on a quantum processor using $p$ repetitions of two Hamiltonians---a problem Hamiltonian $H_{\textrm{P}}$ and a driver Hamiltonian $H_{\textrm{D}}$---for which a quantum circuit can be seen in Fig. \ref{fig:QAOA}. $H_\textrm{P}$ is a Hamiltonian defined by a combinatorial optimisation problem instance that we intend to solve with \begin{equation}
    H_{\textrm{P}} = \sum_{z=0}^{2^n-1} C(z)\ket z\bra z
\end{equation}
where $C$ is the value of the optimisation problem's objective function taking input strings $z$. The driver Hamiltonian in QAOA takes the usual form of 
\begin{equation}
H_{D} = \sum_{i=1}^nX_i
\end{equation}
and is chosen for its ease of implementation as a non-interacting Hamiltonian, whilst still facilitating population transfer between any two given states.
\begin{figure}
    \centering
    \includegraphics[scale =1]{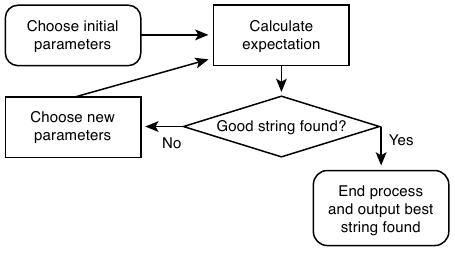}
    \vspace{-.5cm}
    \caption{ The process followed in QAOA. A loop of calculating expectation values and changing parameters is run until a satisfactory string is found.}
    \label{fig:QAOA_flowchart}
\end{figure}
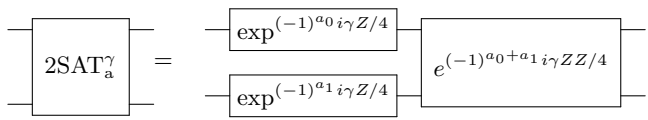
\begin{figure}
\begin{center}\mbox{
\Qcircuit @C=1em @R=0.7cm {
& \multigate{1}{\rm{2SAT}^\gamma_a} & \qw \\
& \ghost{\rm{2SAT}^\gamma_a} & \qw 
} \raisebox{-.5cm}{=} \quad
\Qcircuit @C=1em @R=1em{
 & \gate{\mathrm{e}^{(-1)^{a_0}i\gamma Z / 4}} & \multigate{1}{\mathrm{e}^{(-1)^{a_0 + a_1}i\gamma ZZ /4}} & \qw \\
& \gate{\mathrm{e}^{(-1)^{a_1}i\gamma Z / 4}} & \ghost{\mathrm{e}^{(-1)^{a_0 + a_1}i\gamma ZZ /4}}& \qw } 
}\end{center} 
\vspace{-.3cm}
\caption{Circuit showing the decomposition of a SAT-clause problem Hamiltonian using only $Z$-type operators. $a$ describes the type of the SAT clause with $a=0$ being an OR clause between un-negated variables.} \label{sat}
\end{figure}

To solve a problem with QAOA, the QAOA circuit is run a number of times and the output string measured to calculate an expectation value of $H_{\textrm{P}}$ under the QAOA ansatz state for the current parameters 
\begin{equation}
\langle H_{\textrm{P}} \rangle _{\vec \beta,\vec\gamma} = \bra{\vec \beta,\vec\gamma} H_{\textrm{P}} \ket{\vec \beta,\vec\gamma} \end{equation} 
as in figure  \ref{fig:QAOA}. With or without some post processing \cite{barkoutsos2019improving} this value is handed to a classical optimiser with the aim of producing new parameters via a classical black box optimisation strategy. The expectation value of the problem Hamiltonian is computed again and the process is repeated for either a fixed amount of time or until a satisfactory solution to the problem is discovered. A flowchart depiction of this process is demonstrated in figure \ref{fig:QAOA_flowchart}. Problem-independent success in QAOA is measured in terms of the \textit{mean approximation ratio} defined by:
\begin{equation}
     \frac{\langle H_{\textrm{P}} \rangle _{\vec \beta,\vec\gamma}}{\max_\psi \bra{\psi}H_{\textrm{P}}\ket{\psi}}.
\end{equation}
Combinatorial optimisation problems with clauses encompassing at most two bits can be expressed in terms of two-qubit-$ZZ$ interactions and single-qubit-$Z$ rotations. Problems in which the clauses are local to more bits require higher order terms and are therefore generally out of reach of NISQ quantum computers. Two problems discussed in the literature that do not concern terms of order higher than $2$ are the problems of MAX-CUT and MAX-2-SAT. MAX-CUT, defined on a problem graph in which each vertex is a binary variable, is a problem in which the objective is to find the graph partition such that the number of edges crossing said partition is maximised. The clauses of the problem, or edges of the problem graph are of the type XOR between problem variables. XOR admits the truth table $00,01,10,11 \to 0,1,1,0$ which can be decomposed into a $Z$-based Hamiltonian following theorem 10 of Ref. \cite{hadfield2018quantum}. A MAX-CUT clause, therefore, manifests in the problem Hamiltonian as 
\begin{equation}
H_{\textrm{C,}jk} = \frac{1}{2}\left(I - Z_jZ_k\right) = \textrm{diag}(0,1,1,0).
\end{equation}
The identity in this expression has no effect other than to keep the Hamiltonian non-negative such that the diagonal corresponds to the number of edges a given allocation cuts.

In recent literature, the problems of $2$- and $3$-SAT have seen significant attention due to the presence of \textit{reachability deficits} \cite{akshay2019reachability} in the depth of QAOA required to find an optimal solution. MAX-2-SAT encompasses a more general set of problems than MAX-CUT, with MAX-CUT problems form a subset of possible MAX-2-SAT problems. Two 2-SAT clauses can be combined to construct a CUT clause but a 2-SAT clause cannot be constructed from multiple CUT clauses, since 2-SAT clauses saliently contain single-qubit-$Z$ terms. A 2-SAT clause between two bits can take four forms: $(b_1\vee b_2),(b_1\vee \neg b_2),(\neg b_1\vee b_2),(\neg b_1\vee \neg b_2)$. The logical OR operation $\vee$ yields a truth table $00, 01,10,11 \to 0,1,1,1$ that we can express as a diagonal Hamiltonian $\textrm{diag}(0,1,1,1) = I - \ket{\vec 0}\bra{\vec 0}$. Using the same procedure as before, the four 2-SAT clause types have problem Hamiltonians on the two constituent qubits of \begin{equation}
H_{\vec{a}}= I - \ket{\vec a}\bra{\vec a}
\end{equation} 
which yields a $Z$ operator decomposition as \begin{equation}
H_{\vec{a}} = I - \frac{1}{4}\left[ I + (-1)^{a_0}Z_0 + (-1)^{a_1}Z_1 + (-1)^{a_0 + a_1}Z_0Z_1 \right]
\end{equation}
where $\vec a$ is a binary vector denoting which of the four clause types is used. Figure \ref{sat} shows this decomposition in circuit form.

Both MAX-CUT and MAX-2-SAT are NP-complete problems \cite{karp1972reducibility}, meaning that any other NP-complete problem may be reduced to these problems. Such reductions, however, are unlikely to provide useful implementations on NISQ devices due to large polynomial increases in the number of clauses and variables required to express a reduced problem.

For the size of a MAX-SAT or CUT problem that is of small enough dimension to fit on a near-term quantum computer, a classical home computer can easily solve problems with a brute force approach. With top supercomputers in the world operating in the hundreds of petaflop per second range \cite{top500}, one can roughly estimate that such a computer running for a day would be capable of solving at most a 70-80 bit problem via brute force ($\log_2(0.5\times 10^{18}\times 60^2 \times 24) \approx 75$) This assumes one evaluation of a cost function per floating point operation and is thus a generous upper bound. Competitive SAT solvers do not use brute force methods, but heuristics. An annual SAT solving competition features problems on the order of thousands or tens of thousands of bits \cite{bacchus2020maxsat}. It is yet unknown whether NISQ algorithms will provide competitive heuristic methods for problems in this range of 100-10,000 bits, with the greatest problem size attempted with QAOA being 23 qubits. Such is the nature of heuristic methods that the success or failure of quantum algorithms at this classically difficult scale is best determined via testing real devices at such scales.

\section{Digital Analog QAOA}\label{DAQAOA}

In DA-QAOA we use the DA paradigm to perform a QAOA-approximating algorithm. We take access to the device Hamiltonian
\begin{equation}
H_{\textrm{Device}}(t) = f(t)H_{\rm resource} + \alpha \sum_{i=1}^n\left( x_i(t) X_i +z_i(t) Z_i \right)
\end{equation}
where
\begin{equation}
H_{\rm Resource} = \sum^n_{j<k}r_{jk} Z_jZ_k.
\end{equation}
In the stepwise scheme (sDA-QAOA), we assume control over the parameters $f,x_i,z_i$ each taking values from $\{0,1\}$. In the banged scheme (bDA-QAOA), $f$ is always set to $1$ and only the single-qubit parameters may be altered. The single qubit terms are stronger than the resource Hamiltonian by the factor $\alpha \geq 1$ and in typical applications $\alpha$ is expected to fall between $10-1000$ depending on architecture \cite{ballance2016high,linke2017experimental}. Though current devices tend to exhibit a ratio of single qubit rotation speed to interaction strength at the lower end of this range, they have little to gain from faster single-qubit operations, since they are typically limited by two-qubit interaction times and fidelity. We therefore expect that a device optimising for DA applications could be engineered for greatly higher ratios $\alpha$. During driving in bDAQC, all single-qubit-$X$ operations are set to $1$, $Z$ terms to $0$, giving a driver Hamiltonian of
\begin{equation}
H_{\textrm{bDA-Driver}} = \sum_{j<k} r_{jk}Z_jZ_k + \alpha \sum_{i=0}^nX_i
\end{equation}
applied for device time given by the variational parameter $\beta$ divided by the driver strength $\alpha$ with $\beta \in [0,\pi]$. During the DA resource Hamiltonian steering operations, we use a similar Hamiltonian in which only a specific set of single-qubit-$X$ terms are active. As described in section \ref{section-DASCHEME}, we wish to implement a full $X$-gate before and after each resource block. The time to apply this gate will be $\Delta t = \frac\pi{\alpha}$. Applying the DA-QAOA device Hamiltonian for a single QAOA layer thus effects the following unitary
\begin{equation}
    U_{\rm DA-QAOA} = \mathcal{T} \exp(-i\int_{t = 0}^{t_{\textrm{total}}}  H_{\rm Device} (t)\,\,dt), \label{DAQAOA_UNITARY}
\end{equation}
with $\vec x(t)$ defined by the aforementioned matrix inversion procedure, $\mathcal{T}$ is the time-ordering meta-operator and $\vec z(t)$ used in the case that we are solving a SAT problem. $t_{\rm total}$ is the sum of all times in the non-negative DA time vector multiplied by the variational parameter $\gamma$ in addition to the driving time $\beta / \alpha$. A depiction of this device Hamiltonian used to apply a MAX-CUT problem Hamiltonian is presented in Fig. \ref{fig:compilation}.

\section{Compilation Costs of DA-QAOA}

\begin{figure*}
    \centering
    \includegraphics{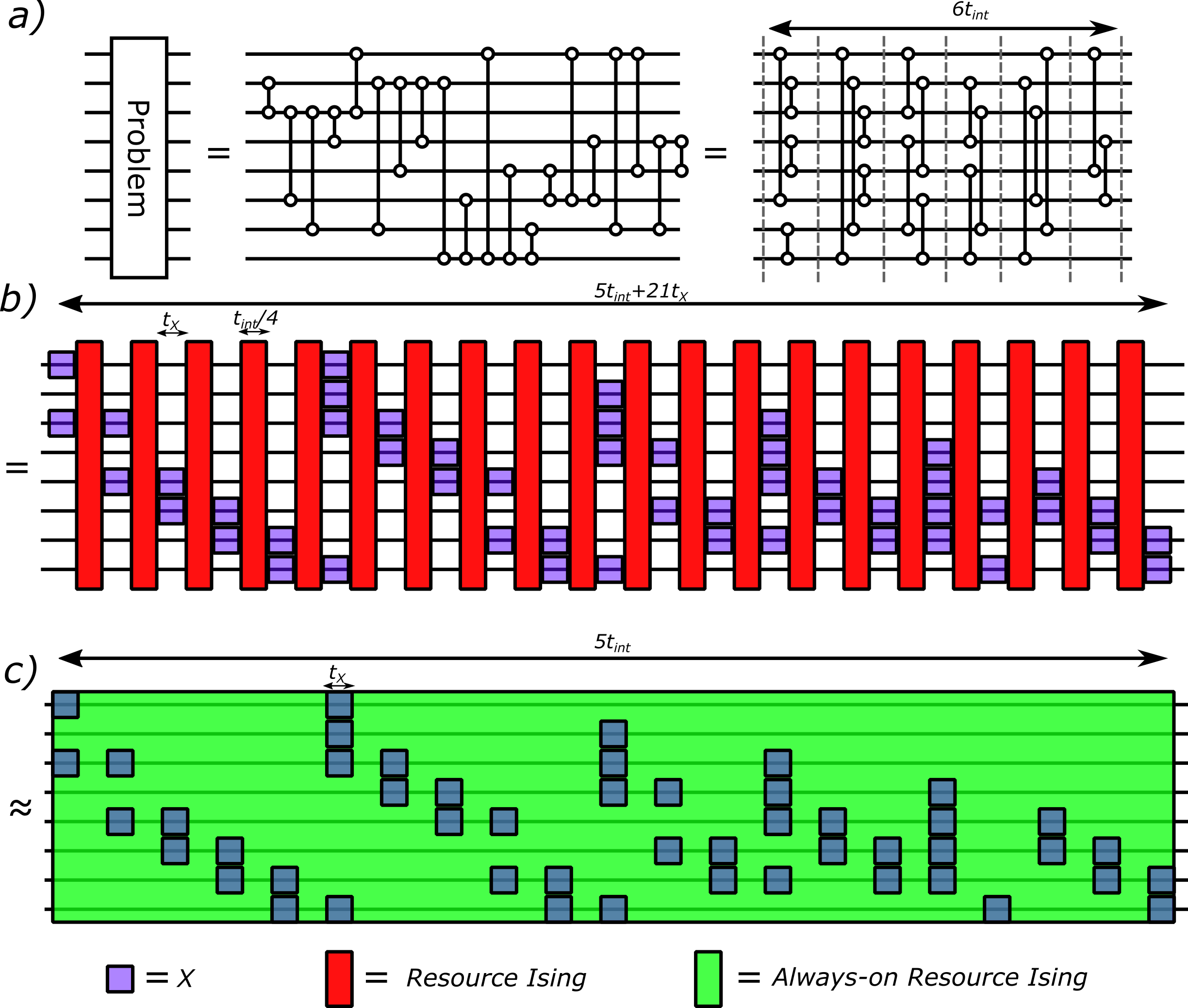}
    \caption{Circuit compilations for an $8$-qubit MAX-CUT problem on a $5$-regular graph. a) A decomposition of the MAX-CUT problem Hamiltonian into $ZZ$ interactions. The lines connected to two open circles represent $ZZ$ interactions applied for time $t_{int}$. On the right hand side, we see that the circuit can be parallelised into six time-steps, each of which sees one qubit interact with only one other qubit at a time. Six time-steps corresponds to the maximum degree of a vertex plus one. b) The same circuit can be compiled into the scheme of sDAQC in which a resource all-to-all homogeneous Ising Hamiltonian is turned on and off, punctuated by single qubit gates. This decomposition requires $20$ uses of the resource Hamiltonian and $50$ single-qubit-$X$ operations. The time taken to apply the problem Hamiltonian is $5t_{int}+21t_{X}$. c) Finally, we compile the problem Hamiltonian in the bDAQC scheme, in which the resource Hamiltonian remains on throughout the procedure. This circuit only approximates the time evolution invoked by the QAOA problem Hamiltonian but can be carried out in time $5t_{int}$ and also with $49$ single-qubit-$X$ operations. In the limit of infinitely fast $X$ gates, c) is equivalent to a) and b).}
    \label{fig:compilation}
\end{figure*}

\begin{figure*}
    \centering
    \hspace{-0.5cm}\includegraphics[scale =0.79]{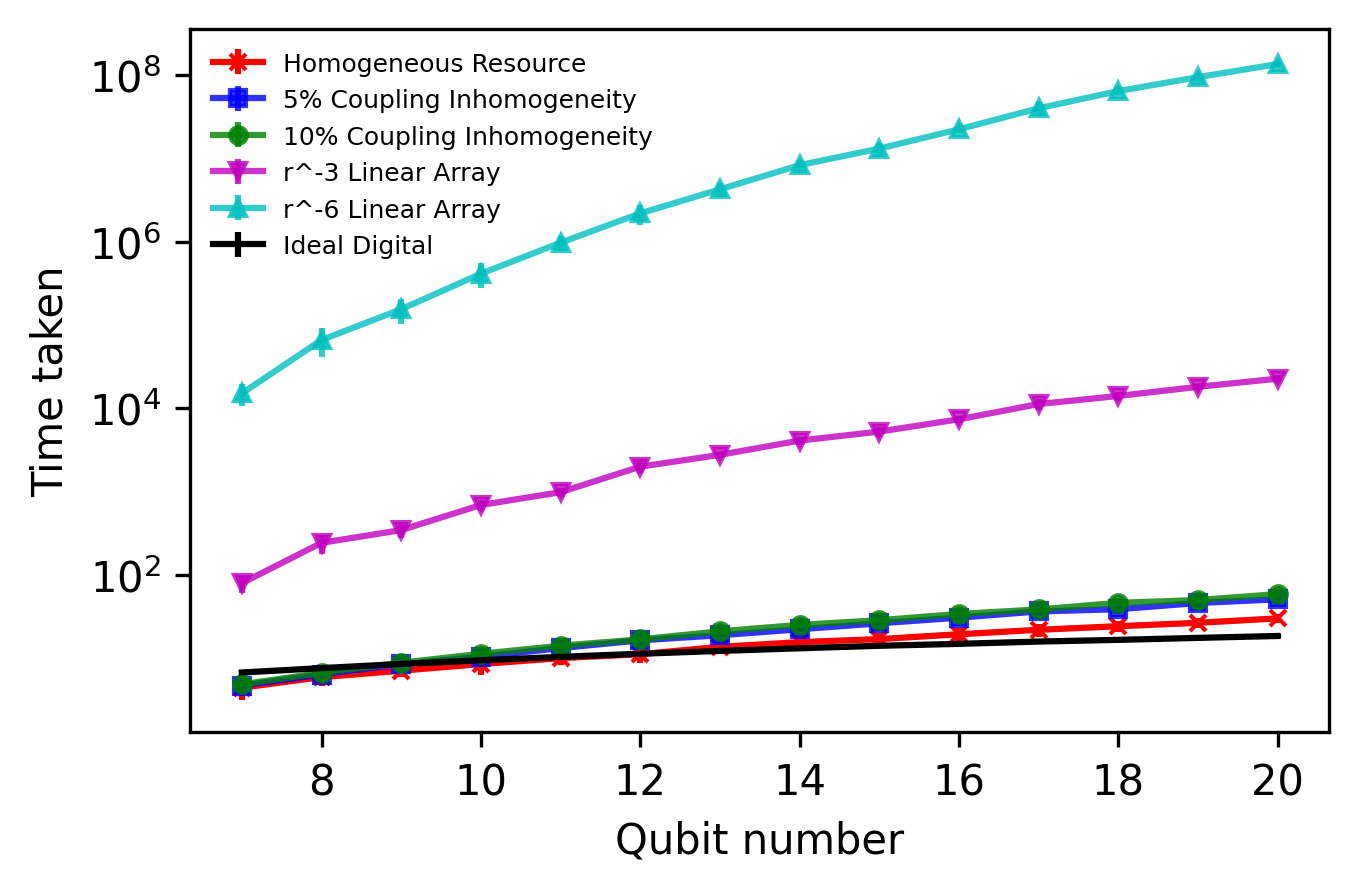}
    \hspace{-0.2cm}\includegraphics[scale =0.79]{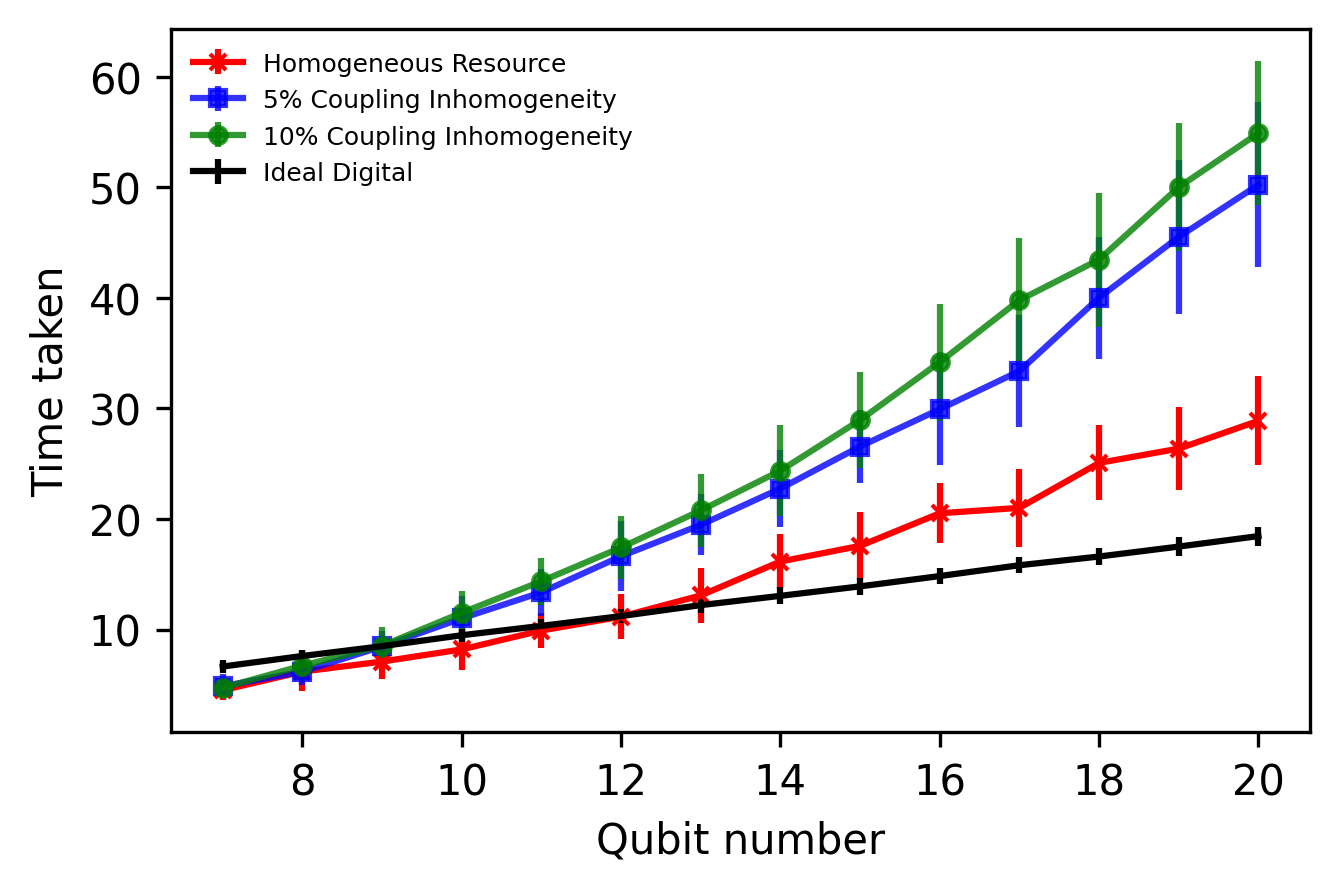}
    \caption{Plots showing the on-device required time for implementation of a QAOA problem Hamiltonian. In this case random Erd\H{o}s R\'enyi MAX-CUT problems are used with a filling factor of $0.75$. Units of time are defined relative to the native device resource interaction strength. In these plots we show the time taken when using various possible resource Hamiltonians. The left hand side demonstrates that if the resource Hamiltonian available varies across orders of magnitude in coupling strength, as happens in the case of an inverse power law coupling between qubits in an array, the time taken in the digital analog scheme becomes extremely large (and likely impractical). The right plot shows the series in the left excluding the inverse power law couplings. A homogeneous resource Hamiltonian is competitive with an idealised digital compiling scheme, though higher for qubit numbers higher than 10. The upper blue (square marker) and green (circle marker) lines demonstrate that if we use a resource Hamiltonian with normally distributed couplings close to $1$ (standard deviations 0.05 and 0.1), the time taken is longer. The significant gap between the Homogeneous and non-homogeneous series occurs as the periodicity of the homogeneous resource Hamiltonian's effected time evolution can be exploited to reduce the idle time. The units for time displayed in the plot are dimensionless multiples of a timescale defined by the resource Hamiltonian.}
    \label{fig:QAOA_DAvsD}
\end{figure*}

In this section the cost in on-device time to perform QAOA using DAQC and different resource Hamiltonians is evaluated. We include in our comparison the time taken by a completely digital quantum computer under reasonable assumptions. We emphasise here that the time taken to perform an algorithm is only a good indication of the fidelity of the algorithm's experimental implementation (or quality of solution) if the device running the experiment is \textit{coherence limited}. In contemporary quantum processors, the limitation is typically not coherence time but the error incurred during the use of two-qubit operations, per operation. An evaluation of whether a device using a DAQC or DQC paradigm performs better would require in-depth knowledge of the error mechanisms of a device operating in the respective paradigm. Such an analysis is expected to favour DAQC given the reduction of errors from turning couplings on and off. 

When comparing the performance of a device making use of the DAQC paradigm to a device running completely digital computations, we must make fair assumptions concerning the capability of each device. We compare the case in which both DAQC and DQC can perform interactions between any pair of qubits. In DQC, the key  limitation we apply---besides the differing error models that are expected to comprise the main advantage of DAQC---is the inability to perform simultaneous two-qubit gates on a single qubit. A given QAOA problem Hamiltonian in DQC must therefore be decomposed into a number of time-steps. This number of time steps for a graph-based problem can be shown to, at most, equal the maximum vertex degree of the problem graph plus one. DAQC in comparison, applies all operations at once, but must utilise many time-blocks to time-average the device resource interaction to the problem Hamiltonian of interest. An example of a QAOA MAX-CUT problem compiled to both the DQC and DAQC is found in figure \ref{fig:compilation} for a $5$-regular MAX-CUT problem on a $8$ qubit device. One notes that for this particular problem, the DA circuit can be performed faster than the digital for sufficiently fast $X$ gates.

Comparisons of the time taken for to implement a problem Hamiltonian are presented in figure \ref{fig:QAOA_DAvsD}. In this plot we compare Hamiltonians from section \ref{hardware}, with homogeneous and inhomogeneous resource interactions. For the inhomogeneous resource interactions, we use couplings $r_{jk}\sim \mathcal{N}\left(1,\delta^2\right)$ where $\delta$ is the fractional standard deviation of the coupling strength. Values of $5\%$ and $10\%$ are used for this inhomogeneity. For the $|r|^{-6}$ and $|r|^{-3}$ power law behaviour, we assume that qubits are placed on in a linear array. For a fair comparison between these resource Hamiltonian and the others, we scale the couplings such that the average coupling between two qubits is the same for all resource Hamiltonians used. Disregarding the speed advantage from exploiting the periodicity of a homogeneous resource, small deviations in the couplings do not greatly affect the compiled time. If, however, any individual coupling becomes especially small, the compilation time grows correspondingly large.

The asymptotic scaling of the algorithm is at worst $O(n^2)$ as $n(n-1)/2$ interaction windows are needed to simulate arbitrary Hamiltonians.

\section{High-Connectivity NISQ Hardware Platforms for DA-QAOA}\label{hardware}

In this section we consider potential hardware realisations of DA-QAOA. A NISQ device able to solve a wide variety of combinatorial optimisation problems running QAOA would require a highly connected quantum device to avoid the need for swapping operations. One could then utilise platforms in which non-local interactions occur natively, while benefiting from the reduced control overhead provided by the digital analog scheme. It has been proposed to use the digital analog scheme to compile \textsc{swap} gates themselves to sequences of digital and analog blocks \cite{galicia2020enhanced}. For realistic near-term hardware, however, we expect any algorithm utilizing swap operations to be out of reach, whether compiled to digital gates or to digital-analog time blocks, due to their excessive contribution to circuit depth and, therefore, decoherence.

Generally considered to be the most mature platform for quantum computing, superconducting solid state qubit architectures tend to have low connectivity due to their 2d-designed nature and are, in current manifestations, not an ideal candidate for performing DA-QAOA \cite{krantz2019quantum}, the potential use of this platform was explored for DAQC in transmon qubits utilizing the cross resonance effect in work by Gonzalez-Raya et. al. \cite{gonzalez2021digital}. Other systems, for example, Rydberg neutral atom arrays or cold, trapped-ion architectures allow for native interactions between all qubits in a device.

Rydberg neutral atoms are atoms in which one or more electrons are in a highly excited state. Excited states of these atoms have high lifetimes owing to their large spatial extent and, therefore, small spatial overlap with the ground state of the atom \cite{schauss2017finite}. Optical latices of Rydberg atoms can feature non-local, all-to-all Van-der-Waals interactions scaling with $|1/r|^6$ for distances $r$ greater than the optical lattice spacing. Such an interaction is highly non-homogeneous but could be utilised for the digital analog scheme.

Trapped ion systems have demonstrated the highest fidelity two-qubit operations \cite{ballance2016high, gaebler2016high} and highest coherence time of any existing platform \cite{wang2017single}. These systems, however, fail to achieve high-fidelity when many qubits are loaded into a trap. This limitation occurs due to frequency crowding of the energy levels used to address the coupling of each ion to the collective motional state of the trapped ions. When the requirement for control over interactions between individual ions in a trap is relaxed, trapped ion platforms perform exceptionally as simulators \cite{zhang2017observation} and qubit numbers competitive with the best superconducting processors can be used to explore physics outside the reach of classical simulation. Interactions between trapped ions scale on the order of $|1/r|^\delta$ with $\delta$ typically varying between $0-3$ \cite{porras2004effective}, with $r$ as the distance between two trapped ions. A system utilising a value of $\delta = 0$---in which the interaction is mediated by the joint vibrational modes---would have a homogeneous coupling if no other non-homogeneous behaviour is present between pairs of ions. The case of $\delta = 3$ occurs when the interaction is mediated purely via spin-spin interactions, incurring a dipolar decay law.
\section{Results}

In this section the main computational results of our work are presented. In the first subsection we present the performance of bDA-QAOA in comparison to a standard QAOA circuit for a set of chosen randomly generated problems. In the second subsection we analyse this further to demonstrate that the algorithm is performing better than one might expect for a banged DA algorithm and that this boost in performance results from the variational freedom of DA-QAOA.

\subsection{Performance of \MakeLowercase{b}DA-QAOA}

\begin{figure*}
    \centering
    \includegraphics[scale =0.92]{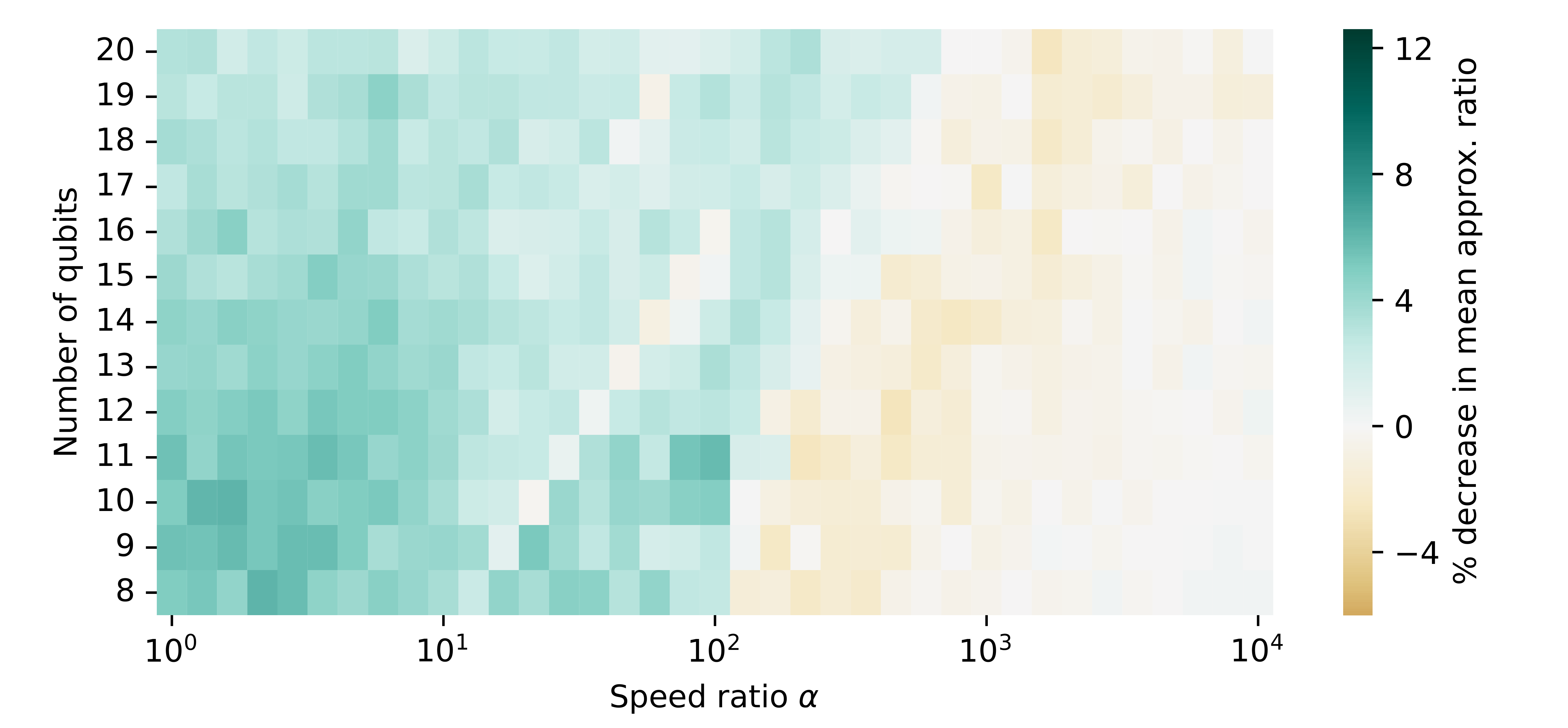} 
    \caption{The percentage difference between the mean approximation ratio attained by bDA-QAOA and error-free QAOA, averaged over $50$ randomly generated MAX-CUT problems with constant filling factor $p_{\rm clause} = 0.7$. Blue (dark region on the far left side of the figure) colours indicate that the bDA-QAOA ansatz state is worse than that provided by error-free QAOA, whereas, shades of brown (the light gray region immediately left of the white region of zero error) indicate an improvement of the bDAQC over error free QAOA. On the $x$-axis, the ratio $\alpha$ of single-qubit to problem Hamiltonian term strength is seen where error-free QAOA exists in the limit as this ratio becomes infinite.}
    \label{fig:QAOA_DA_Approach}
\end{figure*}

In bDA-QAOA we perform QAOA using the ansatz state prepared by applying QAOA layers of the form described in equation (\ref{DAQAOA_UNITARY}) as \begin{equation}
    \ket{\vec \beta,\vec\gamma}^{\alpha,\rm{DA}} = U_{\alpha\textrm{-DA-QAOA}}\ket{+}^{\otimes n}.
\end{equation}
bDA-QAOA introduces errors in the form of the misspecification of the problem and driver Hamiltonians. Between these two, due to differing times taken on device to perform and that the driver is generic to all problems, the misspecification of problem Hamiltonian is likely to introduce more detrimental error. This problem of misspecification is not new to the field of quantum optimisation and is known in quantum annealing literature as $J$-chaos, in which critical characteristics of a problem to be solved are not correctly incorporated into the dynamics of an annealing device. Such issues can be fatal to the performance of adiabatic quantum computing if error mitigation strategies are not utilised \cite{pearson2019analog}. bDA-QAOA finds connection to quantum random walk algorithms \cite{callison2019finding} and adiabatic quantum computing in that the problem Hamiltonian and single-qubit driving operators are performed simultaneously. One might therefore expect that simply running a problem Hamiltonian at the same time as a driver in QAOA should not be fatal, in fact, scheduled quantum random walks and diabatic quantum computing are active fields themselves \cite{morley2019quantum, muthukrishnan2015diabatic}. Simulations performed of bDA-QAOA in which the resource Hamiltonian is identical to the problem Hamiltonian to be solved, such that no DA steering single-qubit operations are required, indeed, showed no discernible net-negative impacts when compared to error-free QAOA. 

Coherent errors occurring in bDA-QAOA with a non-problem-specific resource Hamiltonian, however, are expected to be more damaging than the errors in QAOA with an always-on problem Hamiltonian. These errors will result in a less problem-specific QAOA ansatz state which in turn would be expected to result in a worse expected approximation ratio at a given depth. Figure \ref{fig:QAOA_DA_Approach} displays the mean approximation ratio attained by the bDA-QAOA ansatz state. For high $\alpha$ we see a regime in which, as expected, bDA-QAOA performs identically to error free QAOA. Secondly we see an intermediate regime where minor increases in the mean approximation ratio are observed. Finally, in the case of low $\alpha$ we observe consistently worse results for bDA-QAOA, due to problem-misspecification induced by coherent DAQC errors.

\subsection{Variational Resilience of DA-QAOA to DA Errors}

\begin{figure*}
    \centering
    \includegraphics[scale=0.92]{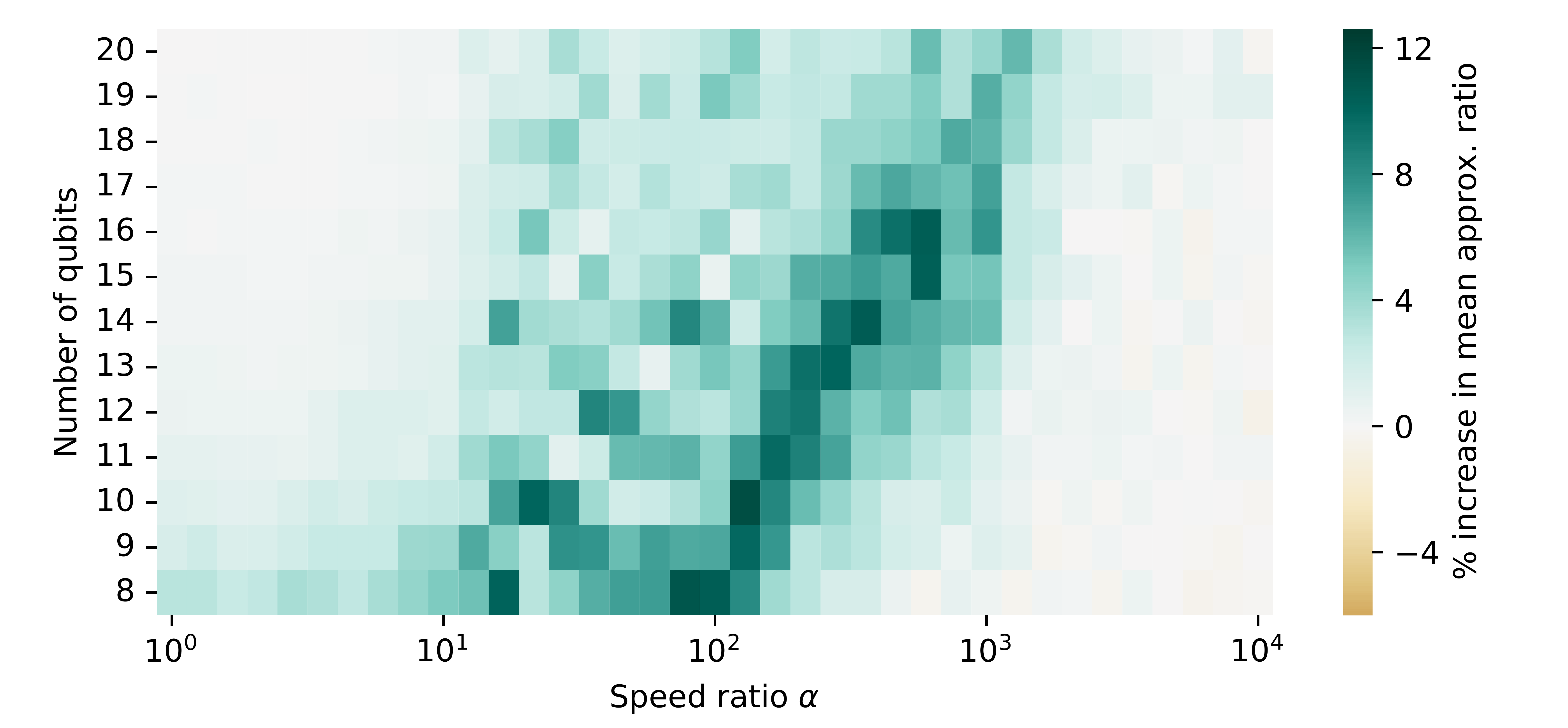}
    \caption{The percentage difference in mean approximation ratio attained by bDA-QAOA at parameters maximising error-free QAOA and bDA-QAOA with optimised parameters, averaged over $50$ randomly generated MAX-CUT problems with constant filling factor $p_{\rm clause} = 0.7$. On the $x$-axis, the ratio $\alpha$ of single-qubit to problem Hamiltonian term strength is seen where error-free QAOA exists in the limit as this ratio becomes infinite. Darker colours imply that for the concerned speed ratio and qubit number, the variational nature of QAOA can account for differences between bDA-QAOA and the ideal algorithm. This plot shows the benefit of using a variational algorithm such as QAOA over a non-variational algorithm in the DA context where coherent error is introduced.}
    \label{fig:hungry_optimiser}
\end{figure*}

QAOA is a variational algorithm. It is expected that variational algorithms have better error tolerance properties due to the fact that a classical optimiser can account for systematic coherent over- or under-rotations and other systematic coherent errors \cite{mcclean2016theory}, making variational quantum algorithms appealing candidates for NISQ quantum computing. QAOA works by finding a parameter set $\vec \beta^*,\vec\gamma^*$ maximising $\langle H_{\textrm{P}} \rangle _{\vec \beta,\vec\gamma}$. However, when we change the QAOA ansatz operators to those of bDAQC, there is no clear reason why the parameters $\vec \beta^*,\vec\gamma^*$ maximising $\langle H_{\textrm{P}} \rangle _{\vec \beta,\vec\gamma}$ also maximise $\langle H_{\textrm{P}} \rangle _{\vec \beta,\vec\gamma}^{\alpha,\rm{DA}}$ where 
\begin{equation}
\langle H_{\textrm{P}} \rangle^{\alpha,\rm{DA}}_{\vec \beta,\vec\gamma} = \bra{\vec \beta,\vec\gamma}^{\alpha,\rm{DA}} H_{\textrm{P}} \ket{\vec \beta,\vec\gamma}^{\alpha,\rm{DA}}.
\end{equation}
 Figure \ref{fig:hungry_optimiser} suggests that this is not the case and shows that significant increase in the success probability of QAOA result from the variational freedom of the algorithm. Figure \ref{fig:hungry_optimiser} should be understood to demonstrate the parameter regimes for which it makes more sense to perform a variational algorithm such as QAOA rather than a fixed gate sequence algorithm such as the the quantum Fourier transform. For high $\alpha$, the error introduced by the scheme is negligible, and both variational algorithms and fixed sequence algorithms will perform similarly. In the middle of the plot, a dark turquoise band can be seen implying that while a non-variational algorithm will have low fidelity due to the presence of DA-induced coherent errors, the variational algorithm still functions. For low enough $\alpha$, we enter a regime in which even the variational algorithm fails to recover any performance through altering parameters. We interpret that this lack of ability of DA-QAOA to absorb error in the low $\alpha$ regime is a manifestation of barren plateaus in the objective function \cite{mcclean2018barren}. Barren plateaus are a feature discovered to occur in the optimisation landscapes of quantum neural networks. When parameterised random circuits are used as ansatze in variational algorithms, the gradient of the objective function with respect to the variational parameters is observed to become exponentially small in the number of qubits used. When $\alpha$ reduces to a certain value, we interpret that the DA-QAOA ansatz loses specificity to the problem Hamiltonian of interest. The variational form used for optimisation no longer bears similarity to the objective function used and is, consequently, no better an ansatz than a random quantum circuit. At this point of low $\alpha$, we observe that the gradient of our objective function with respect to the variational parameters $\vec \gamma, \vec \beta$ tends to become prohibitively small and the approximation ratio attained therefore varies little with differing parameters.



\section{Analytical Fidelity Bounds for \MakeLowercase{b}DA-QAOA}

In this section we demonstrate that the error introduced by performing QAOA in the banged digital analog paradigm in comparison to regular QAOA can be analytically bounded. In particular, we  place a lower bound on the fidelity of a state that arises from a bDA-QAOA circuit in comparison to a state prepared by error free QAOA. This error consists of multiple steps, each of which is of the same nature as that occurring when Trotterizing a Hamiltonian with non-commuting terms for simulation. In the case of bDAQC induced error there are, however, two complications. Firstly, there is only a single Trotterization time-step which cannot be made arbitrarily small with the use of higher numbers of Trotter blocks. Secondly, we use Trotterization in reverse in this fidelity bound. In a usual Trotterization procedure, the simultaneous case is `correct' and the digitalised version introduces error. In the digital analog scheme, however, the opposite is true. The sequential Hamiltonian is ideal and the simultaneous Hamiltonian introduces error. These differences do not affect the validity of the bound, since the bound used in this work is valid for arbitrarily large time-steps. The bound used, to our knowledge, represents the current lowest bound on Trotter error \cite{kivlichan2019improved} and limits the size of the greatest eigenvalue of an operator derived via the difference of two unitaries.   The first unitary is that generated by sequential application of two Hamiltonians $A,B$, with the second generated by simultaneous application of such Hamiltonians:
\begin{multline}
    \big\|\exp(iA/2)\exp(iB)\exp(iA/2) - \exp(i(A+B))\big\|  \\ 
    \leq\frac{1}{12} \big\|\left[\left[A,B\right],B\right] \big\| + \frac{1}{24} \big\|\left[\left[A,B\right],A\right]\big\|.
\end{multline}
The system Hamiltonian during time periods in which both the single-qubit operations and the resource Hamiltonian are active is
\begin{equation}
H_{\textrm{Steering, }\mu} = \alpha\sum_{m\in S_{\mu}}X_m + H_{\textrm{Resource}}
\end{equation}
where $S_{\mu}$ is the index set of $X$ operators applied following DA time-block $\mu$. There are $n(n+1)/2 + 2$ periods of time for which we apply this bound, $n(n+1)/2$ sets of single-qubit operations following interaction blocks, one idle block and the driving block of the QAOA algorithm. We wish to compute the error resulting from using this, rather than its single-step Trotterization. Of these error effecting blocks, $n-3$ will have four full single-$X$ rotations, $n(n-1)/2 - (n - 3) + 1$ will have two full single-$X$ rotations and one, the driver, will have $n$ single-$X$ rotations of duration $\beta \leq \pi$. 
We can allocate $A = t H_{\textrm{R}}$ and $B = t\alpha\sum_{m\in S_{\mu}}X_m $. Where every term in $H_{\textrm{R}}$ consists only of Pauli-$Z$ strings. We can thus write
\begin{multline}
    \left|\left|e^{\frac{i tH_{\textrm{R}}}{2}}e^{it\alpha\sum_{m\in S_{\mu}}X_m}e^{\frac{i tH_{\textrm{R}}}{2}} - e^{it\left(\alpha\sum_{m\in S_{\mu}}X_m + H_{\textrm{R}}\right)}\right|\right|\\ \leq
    \frac{\alpha^2t^3}{12} \left|\left|\left[\left[ H_{\textrm{R}},\sum_{m\in S_{\mu}}X_m\right],\sum_{m\in S_{\mu}}X_m\right]\right|\right| \\+ \frac{\alpha t^3}{24} \left|\left|\left[\left[ H_{\textrm{R}},\sum_{m\in S_{\mu}}X_m\right], H_{\textrm{R}}\right]\right|\right|
    \label{fullbound}
\end{multline}
because every block in the DA-QAOA setting will be surrounded by resource blocks, it does not matter whether the Trotterization is symmetric or asymmetric. To calculate the first commutator, we can expand the sums and compute each individual term
\begin{multline}
    \sum_{m\in S_{\mu}}\sum_{m'\in S_{\mu}}\sum_{j<k}\left[\left[ Z_jZ_k,X_m\right],X_{m'}\right]\\
    =\sum_{m\in S_{\mu}}\sum_{m'\in S_{\mu}}\sum_{k>m}2\left[ Z_mZ_kX_m,X_{m'}\right]\\
    +\sum_{m\in S_{\mu}}\sum_{m'\in S_{\mu}}\sum_{j<m}2\left[ Z_jZ_mX_m,X_{m'}\right]\\
    =\sum_{m\in S_{\mu}}\sum_{m'\in S_{\mu}}\sum_{j \neq m}2\left[ Z_jZ_mX_m,X_{m'}\right]\\
    =\sum_{m\in S_{\mu}}\left(\sum_{j \neq m}4Z_mZ_{j}
    +\sum_{m'\in S_{\mu}|m'\neq m}4i Y_mY_{m'}\right).
\label{opsum1}
\end{multline}
We have here assumed that the resource is homogeneous. In a given single-qubit block there will be $s$ possible indices in the corresponding set $S_\mu$. As such we will have $s(n-1)$ $Z$ strings and $s(s-1)$ $Y$ strings. The eigenvalues of sets of terms that can be simultaneously diagonalised will add linearly. We have two orthogonal bases in which eigenvalues are added as such, the sums of which will add in quadrature. The greatest eigenvalue of the entire sum in equation (\ref{opsum1}) can then be written $s\sqrt{(s-1)^2 + (n-1)^2}$. The contribution to the bound in equation (\ref{fullbound}) from this commutator can therefore be written
\begin{equation}
\frac{\alpha^2t^3s\sqrt{(s-1)^2 + (n-1)^2}}{3}.
\end{equation}
For a homogeneous resource the commutator in the second term can be written as 
\begin{multline}
    \sum_{m\in S_\mu}\sum_{j,j'<k,k'}\left[\left[ Z_jZ_k,X_m\right], Z_{j'}Z_{j'}\right] \\
    = \sum_{m\in S_\mu}\sum_{j,j'<k,k'} 2(\delta_{jm} + \delta_{km})\left[ Z_jZ_kX_m, Z_{j'}Z_{k'}\right] \\
    = \sum_{j'<k'}\sum_{m\in S_\mu}\sum_{j \neq m} 2Z_j\left[ Z_mX_m, Z_{j'}Z_{k'}\right]\\
    = \sum_{m\in S_\mu}\sum_{j \neq m}\sum_{j'\neq m} 4 Z_jZ_{j'}Z_mX_mZ_{m} \\
    = \sum_{m\in S_\mu}\sum_{j \neq m}\sum_{j'\neq m} -4 Z_jZ_{j'}X_m,
\end{multline}
where the negative sign is of no consequence. We have $(n-1)^2$ terms per single-qubit-$X$ operator and $s$ $X$-operators giving $s(n-1)^2$ strings. At worst the greatest eigenvalue of this operator sum will be equal to the number of Pauli strings. We therefore obtain a full bound of 
\begin{equation}
\Delta_\mu = \frac{\alpha s t^3}{3} \left(\frac{(n-1)^2}{2} + \alpha\sqrt{(s-1)^2 + (n-1)^2}\right).
\end{equation}
We wish to bound the minimum fidelity of a coherent erroneous operation caused by using bDA-QAOA
\begin{equation}
    f_{\alpha\rm -DA-QAOA} = \min_\psi\left\|\bra\psi U_{\textrm{QAOA}}^\dagger U_{\textrm{$\alpha$-DA-QAOA}}\ket\psi \right\|^2.
\end{equation}
 If the magnitude of the greatest eigenvalue of the difference between two unitaries  operators is bounded as in the Trotterization bound
 \begin{equation}
     \|U- U_{\alpha} \| \leq \Delta
 \end{equation}
then 
\begin{equation}
\|U^\dagger U_{\alpha} -  I\| \leq \Delta
\end{equation}
which yields a bound of
\begin{equation}
\|e^{i|\theta|_\textrm{max}} - 1\| \leq \Delta
\end{equation} 
where $e^{i|\theta|_\textrm{max}}$ is the greatest eigenvalue of $U^\dagger U_{\alpha}$, assuming the eigenvalues are small such that all angles lie on the interval $[-\pi/2,\pi/2]$. The greatest phase acquired under the erroneous evolution can then be related to the greatest eigenvalue bound as 
\begin{equation}
2\sin{\left(\frac{|\theta|_{\rm max}}{2}\right)} = \Delta, \quad \theta = 2\sin^{-1}{\left(\frac{\Delta}{2}\right)}
\end{equation}
where $0 <\Delta <1$. Consider the fidelity of a state under an erroneous operator $ \hat O = U^\dagger U_{\rm imperfect}$ corresponding to the time-step $\mu$ during which a round of single-qubit-operations are performed: \begin{equation}
    \left\|\bra\psi \hat O \ket\psi \right\|^2.
\end{equation}
We can write this in the a basis diagonalising $\hat O$, 
\begin{equation}
\left\|\bra{\psi'} \textrm{diag}\left(e^{i\vec\theta}\right) \ket{\psi'} \right\|^2
\end{equation}
and this fidelity is minimised when the state is an equal superposition of the most positive and most negative argument eigenstates of $\hat O$.
\begin{multline}
\min_{\psi'} \left\|\bra{\psi'} \textrm{diag}\left(e^{i\vec\theta}\right) \ket{\psi'} \right\|^2 \\= \frac14 \left\|\left(\bra{\theta_{\textrm{max}}}+\bra{\theta_{\textrm{min}}}\right) \hat O \left(\ket{\theta_{\textrm{max}}} + \ket{\theta_{\textrm{min}}}\right) \right\|^2 \\
=\frac14 \left\| \bra{\theta_{\textrm{max}}}e^{i\theta_{\textrm{max}}}\ket{\theta_{\textrm{max}}} + \bra{\theta_{\textrm{min}}}e^{i\theta_{\textrm{min}}}\ket{\theta_{\textrm{min}}} \right\|^2 \\
\geq \frac14 \left| e^{i|\theta|_{\textrm{max}}} + e^{-i|\theta|_{\textrm{max}}}  \right|^2 \\
= \cos{(|\theta|_{\max}})^2 
\end{multline}
So the fidelity of a single qubit block $\mu$ is
\begin{equation}
f_\mu \geq \cos{(|\theta|_{\textrm{max}})}^2, \end{equation}
or in terms of the bound of the greatest eigenvalue of this set of single-qubit-operations $\Delta_\mu$ 
\begin{equation}
f_\mu \geq \cos{\left(2\sin^{-1}\left(\frac{\Delta_\mu}{2}\right)\right)}^2 \geq 1-\Delta_\mu^2
\end{equation} 
 with equality in the limit of small $\Delta$.
Using the subadditivity of infidelity \cite{nielsen2002quantum} we can finally express 
\begin{multline}
    f_{\alpha\rm -DA-QAOA} \geq 1- \sum_{\mu=1}^ {n(n-1)/2 + 2} (1-f_\mu)\\
    = 1- \sum_{\mu=1}^ {n(n-1)/2 + 2}\Delta_\mu^2.
\end{multline}
For one set of single-qubit-rotations, the driver, all $X$ terms are active, giving $s = n$. The remaining time-blocks either have $2$ or $4$ single-qubit gates active determined by whether cancellations occur. No cancellations occur for $n-3$ blocks resulting in $s=4$, with the remaining $n(n-1)/2 - (n-2)$ blocks taking $s = 2$. The time taken to perform each block is $t = \pi / \alpha$. For large $n$ we find that the speed of single-qubit gates must increase approximately as the number of qubits squared for high-fidelity with the ideal QAOA state.  While this discussion has considered only the use of a homogeneous resource Hamiltonian in the interest of brevity, this is not critical to the calculation of the bound. A heterogeneous resource would result in a sum of Pauli strings with non-unit pre-factors, which could be subsequently summed straightforwardly, as in the homogeneous case.


\section{Sensitivity to other errors}

Next to the errors discussed in the previous sections that are imminent to the hardware simplification provided by our digital-analog approach, the algorithm is exposed to other sources of errors common to NISQ computing. As detailed error budgets of concrete hardware are currently hard to determine, we would like to qualitatively evaluate their impact on our technique. A detailed evaluation of the effects of differing noise sources applicable to digital analog quantum computing in comparison to a gate based approach was performed in \cite{garcia2021noise}, with results in favor of the digital analog paradigm.

On the one hand, single-qubit gate errors induced by decoherence measured by $T_{1/2}$ will have full impact on this algorithm as these are repeatedly executed. Small errors of the rotation axis will also have full impact as they can be mistaken for a modified problem Hamiltonian. Errors of the rotation angle can be expected to be less critical as some of them can be accommodated in the classical optimisation process. So all in all, single-qubit errors have the same if somewhat smaller impact than in a compiled gate model QAOA. 

Two-qubit gates do not appear directly in our scheme thus avoiding two-qubit gate control errors as well as the additional entry points for noise through fast two-qubit control ports. However, the interaction mediated by the problem Hamiltonian can still create entangled states, which decay faster than non-entangled states. Notably, an $n$-qubit GHZ state dephases in a time $T_2/n$ \cite{dur2004stability}. The precise degree of entanglement needed for a specific problem instance is currently unknown for any quantum optimisation algorithms. Yet, we can summarise that the sensitivity of digital-analog QAOA to two-qubit errors is lower than the compiled version. Given a single qubit error rate, alongside the total execution time of the algorithm relative to $T_2$, the depth at which this algorithm can be faithfully executed could be inferred.

In this estimate we need to keep in mind whether coherent over-rotation errors have an effect different to incoherent errors. This case could occur if they interfered in a structured way. Given the randomisation effect of the problem Hamiltonian to any state, this is unlikely and we expect that their impact is faithfully represented by the measured fidelity. 

\section{Conclusion}

The possibilities of using models of quantum computation less conventional than the standard gate based approach have not been fully considered. In this work, we show that while an alternative approach---the digital analog paradigm---might introduce errors of its own, the device complexity required to control the time evolution of the system can be reduced and errors introduced are of a nature that can be non-fatal to variational algorithms such as QAOA in certain regimes. We demonstrate that the digital analog paradigm is an ideal setting in which to do QAOA, as each problem Hamiltonian operator can be performed in a single DAQC block, that resource Hamiltonians expected from hardware can be utilised to implement QAOA Hamiltonians mitigating swapping overhead associated with mainstream approaches, and that QAOA displays error resilience beyond that of pre-programmed algorithms in the digital analog paradigm. This work presents new possibilities for the design of NISQ devices for combinatorial optimisation, bridging the gap between current devices and full, fault-tolerant quantum computers, bringing hardware closer to the point of demonstrating a quantum advantage for real-world problems.

\section*{Acknowledgments}

The authors acknowledge funding, support and computational resources from Mercedes-Benz AG, DLR, the Basque Government QUANTEK project from the ELKARTEK program (KK-2021/00070) and IT1470-22, Spanish Ram\'on y Cajal Grant RYC-2020-030503-I and the project grant PID2021-125823NA-I00 funded by MCIN/AEI/10.13039/501100011033 and by ``ERDF A way of making Europe'' and ``ERDF Invest in your Future'', as well as from OpenSuperQ (820363) and QMiCS (820505) of the EU Flagship on Quantum Technologies, and the EU FET-Open projects Quromorphic (828826) and EPIQUS (899368). We acknowledge useful conversations with Markus Leder, Tyler Takeshita and Tobias Stollenwerk.




\bibliography{ref} 
\bibliographystyle{apsrev4-1}

\end{document}